\documentclass[review]{elsarticle}

\usepackage{amsmath}
\usepackage{hhline}
\usepackage{mathrsfs}
\usepackage{amssymb}            
\usepackage[english]{babel}
\usepackage[T1]{fontenc}
\usepackage[latin1]{inputenc}
\usepackage{rotating}

\journal{Comm. Nonlinear Sci. Numer. Simulat.}

\begin{document}

\begin{frontmatter}

\title{Design and performance of low-energy orbits for the exploration of Enceladus} 

\cortext[mycorrespondingauthor]{Corresponding author}
\author[UAE]{Elena Fantino\corref{mycorrespondingauthor}}
\ead{elena.fantino@ku.ac.ae}
\author[UAE]{Francisco Salazar}
\ead{francisco.salazar@ku.ac.ae}
\author[IMATI,IFAC]{Elisa Maria Alessi}
\ead{em.alessi@mi.imati.cnr.it}

\address[UAE]{Department of Aerospace Engineering, Khalifa University of Science and Technology, P.O. Box 127788, Abu Dhabi, United Arab Emirates}
\address[IMATI]{Istituto di Matematica Applicata e Tecnologie Informatiche  "Enrico Magenes", Consiglio Nazionale delle Ricerche, Via Alfonso Corti 12, 20133 Milano, Italy}
\address[IFAC]{Istituto di Fisica Applicata "Nello Carrara", Consiglio Nazionale delle Ricerche, Via Madonna del Piano 10, 50019 Sesto Fiorentino (FI), Italy}

\begin{abstract}
The icy moons are in the focus of the exploration plans of the leading space agencies because of the indications of water-based life and geological activity observed in a number of these objects. In particular, the presence of geyser-like jets of water near Enceladus' south pole has turned this moon of Saturn into a priority candidate to search for life and habitability features. This investigation proposes a set of trajectories between Halo orbits about Lagrangian points $L_1$ and $L_2$ in the Saturn-Enceladus Circular Restricted Three-Body Problem as science orbits for a future {\it in situ} mission at Enceladus.
The design methodology is presented, followed by the analysis of the observational performance of the solutions. The conclusion is that the proposed orbits exhibit suitable features for their use in the scientific exploration of Enceladus, i.e., long transfer times, low altitudes, wide surface visibility windows and long times of overflight. 
\end{abstract}


\end{frontmatter}


\section{Introduction}
In the framework of the new generation of solar system exploration missions, high priority is given to the observation of the so-called Inner Larger Moons of Saturn, namely, Mimas, Enceladus, Tethys and Dione \cite{Vision:2011}. In particular, the geyser-like jets venting water vapor, ammonia, salts, hydrogen and organics observed by Cassini at the south pole of Enceladus in 2005, 2008 and 2015 \cite{Porco:2006, Spencer:2006, Parkinson:2007, McKay:2008, McKay:2014, MacKenzie:2016, Khawaja:2019} have placed this moon among the targets to search for life and habitability features in the outer solar system (see also \cite{Hartogh:2011}). The scientific questions related to the nature of Enceladus' ejecta can only be answered by carrying out dedicated missions capable of extended observations of the key features observed during Cassini's close passages. This raises the need for specialised orbits offering long close-up views of the surface of this moon.

The design of science orbits around planetary satellites brings challenges because of the perturbing effect of the planet's gravity. Near-polar orbits around Enceladus are unstable and can only be reached by expensive change-of-plane maneuvers \cite{Scheeres:2001}. Previous studies for planetary probes have managed to identify long-term stable orbits in Saturn-Enceladus Hill's model. For example, Russell \& Lara \cite{Russell:2009a} performed a global grid search in the unaveraged Saturn-Enceladus Hill's model including the spherical lunar gravity terms, and identified long-term stable orbits with altitudes near 200 km and inclinations approaching 65$^{\circ}$. Lara et al. \cite{Lara:2010} computed a higher-order approximation in the averaged Hill's model and applied it to the Saturn-Enceladus system, finding a stable, quasi-circular frozen orbit around the moon with average altitude and inclination of 183 km and 61$^{\circ}$, respectively.

This contribution presents a systematic design and analysis of orbits around Enceladus in the Saturn-Enceladus Circular Restricted Three-Body Problem (CR3BP). This model accounts for the perturbing effect of the planet's gravity and offers very good approximations to $n$-body solutions since the influence of the other moons is negligible owing to their small masses and their large distances from Enceladus.
Since the goal is to observe the south pole of this moon, the orbits must develop in 3D. 
Halo orbits, a type of periodic Libration Point Orbits (LPOs) around  the Lagrange points $L_1$ and $L_2$ \cite{Richardson:1980, Howell:1984} are here computed for the Saturn-Enceladus system and employed as departure and end points of transfer trajectories. These connections exhibit a significant out-of-plane component (inherited by their progenitor Halo orbits) and make close approaches to the surface of the moon. Since they shadow heteroclinic transfers, they will be referred to as s-heteroclinics. 
The computation is based on finding the intersections between the stable and unstable hyperbolic invariant manifolds (HIMs) of the departure and arrival orbits. The reader is referred to the fundamental work of G\'omez et al. \cite{Gomez:2004}, Canalias \& Masdemont \cite{Canalias:2006} and Barrab\'es et al. \cite{Barrabes:2009} for the computation of heteroclinic connections between LPOs in the Sun-Earth and Earth-Moon systems using HIMs, and to Davis et al. \cite{Davis:2018} for the application to the design of a connection between Halo orbits of  $L_1$ and $L_2$ in the Saturn-Enceladus system.
The present investigation confirms and extends the latter work and identifies more connections.
The existence of maneuver-free transfers and the periodic character of the Halo orbits can be exploited to construct a fuel-efficient exploration tour of the moon made of chains of itineraries in which the departure and arrival Halo orbits are used as parking orbits between consecutive transfers and as gates to reach other moons in the system. The latter concept would extend to three dimensions the low-energy, low-thrust inter-moon connections designed in previous contributions by these and other authors \cite{Russell:2009, Fantino:2016, Fantino:2019}.

The s-heteroclinics between Halo orbits at Enceladus are here proposed as science orbits for the observation of the surface features of interest. Hence, an important aspect of the work is the study of the observational performance of these trajectories. Kinematical and geometrical parameters such as transfer times, distances from the surface, speeds relative to an Enceladus-centered inertial frame, times of overflight, instantaneous and cumulative surface coverage parameters and ground tracks are computed and analysed aiming to assess the suitability of the computed solutions for scientific use.  A preliminary stage of this work was presented in Salazar et al. \cite{Salazar:2019}. 


The paper is organised as follows. Section~\ref{sec:model} summarizes the relevant characteristics of the CR3BP and illustrates the families of Halo orbits around $L_1$ and $L_2$ computed and employed in this work. The methodology adopted for the design of connections between these orbits and the resulting solutions are presented in Sect.~\ref{sec:heteroclinics}. Section~\ref{sec:observ} exposes the observational properties of the computed trajectories. Discussion and conclusions follow in Sect.~\ref{sec:concl}.

\section{The CR3BP, Halo orbits and their stable and unstable HIMs}
\label{sec:model}
\begin{figure}[h!]
\centering
\includegraphics[scale=0.3]{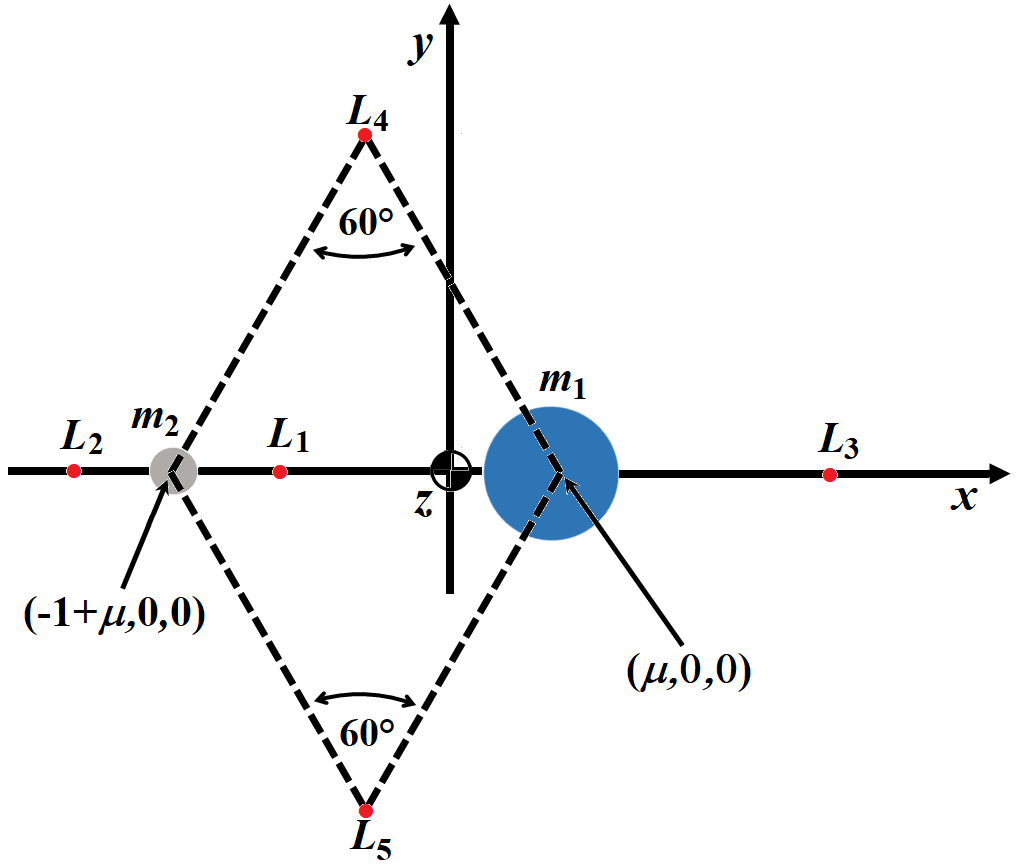}
\caption{The primaries, the third body and the five libration points in the synodical barycentric reference frame of the CR3BP.}
\label{fig:CR3BP}
\end{figure}

The CR3BP models the motion of a massless body (here the spacecraft, S/C) subjected to the gravitational attraction of two primaries of mass $m_1$ (the first primary, here the planet) and $m_2$ (the second primary, here the moon), assumed to move on circular orbits about their center of mass \cite{Szebehely:1967}. The equations of motion of the third body are conveniently expressed in the synodical barycentric reference frame ($O$, $x$, $y$, $z$) with the two primaries stationary on the $x$-axis. The total mass $\left(m_1+m_2\right)$ of the system and the distance $r_0$ between the primaries are adopted as the units of mass and length, respectively. The unit of time is defined by setting the period $T$ of the orbits of the primaries equal to $2\pi$. This corresponds to assigning unit value to their orbital mean motion. As a result, the positions of the primaries in normalised units are $\left(\mu,0,0\right)$ and $\left(\mu-1,0,0\right)$, respectively (Fig.~\ref{fig:CR3BP}), $\mu$ being the normalised mass of the second primary and the mass ratio of the system: $\mu= m_2/ \left(m_1+m_2\right)$. 

For the equations of motion of the S/C in this model, the definition of the Jacobi constant $C_J$ and the existence, location and properties of the five equilibrium points $L_i$ ($i$=1,2,..,5), the reader is referred to fundamental literature, such as \cite{Szebehely:1967}.
When $m_2 \ll m_1$, $L_1$ and $L_2$ (see Fig.~\ref{fig:CR3BP}) approximately lie at the intersections of the $x$-axis with the Hill sphere, centered at the smaller primary and having radius \cite{Capderou:2005}
\begin{equation}
\label{eq:rH}
r_H =  \left(\frac{m_2}{3m_1}\right)^{1/3} r_0. 
\end{equation}
Table~\ref{tab:CR3BP} reports the basic features of the Saturn-Enceladus CR3BP. 
\begin{table}[h!]
\begin{center}
\caption{Basic features of the Saturn-Enceladus CR3BP: mass of Saturn ($m_1$), mass ratio ($\mu$), mean physical radius of Enceladus ($R$), distance between Saturn and Enceladus ($r_0$), orbital period of the system ($T$), radius of the Hill sphere (raw data from \cite{Horizon}).
\vspace{5mm}} 
\label{tab:CR3BP}
\begin{tabular}{cccccc} 
\hline\noalign{\smallskip} 
$m_1$ & $\mu$ & $R$ & $r_0$  & $T$ & $r_H$\\ 
(kg) &  & (km)  & (km) & (day) & (km)\\ 
\noalign{\smallskip}\hline\noalign{\smallskip} 
$5.68336\cdot10^{26}$  & $1.899309 \cdot 10^{-7}$   &  252.1  & $2.38042 \cdot 10^{5}$   & 1.37 & 948.7\\
\noalign{\smallskip}\hline
\end{tabular}
\end{center}
\end{table}

\begin{figure}[h!]
\centering
\includegraphics[scale = 0.170]{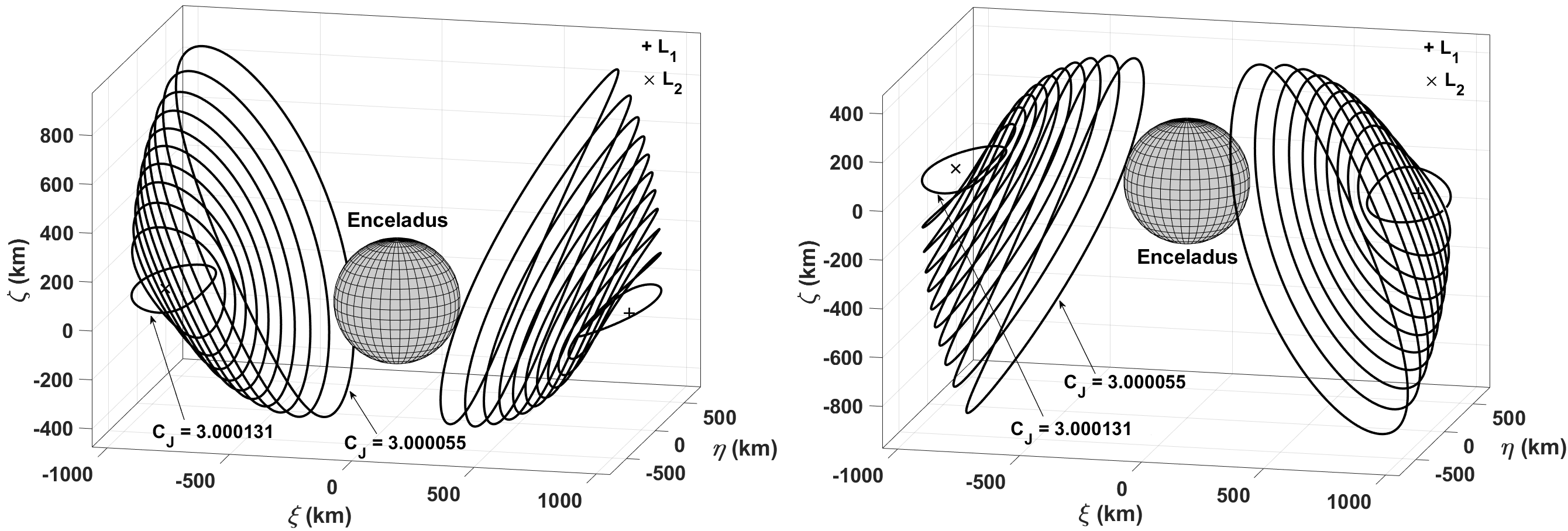}
\caption{Families of Northern (left) and Southern (right) Halo orbits around $L_1$ and $L_2$ (Enceladus-centered synodical frame with unnormalised units).}
\label{fig:Halos}
\end{figure}

The linear approximation of the equations of motion close to an equilibrium point leads to families of LPOs. Halo orbits around $L_1$ and $L_2$ have been computed and used in this work owing to their out-of-plane component, which offers opportunities for extended coverage of Enceladus' surface. Each libration point admits two symmetric families of Halos, the so-called Northern and Southern Halo orbits \cite{Howell:1984}, the symmetry being across the $xy$-plane: a Southern Halo orbit can be obtained from a Northern Halo orbit through the transformation $z \rightarrow -z$, $\dot{z} \rightarrow -\dot{z}$.
Families of Halo orbits have been computed over a wide energy interval and using an energy discretization such that all the families have 100 members at identical values of $C_J$ between 3.000055 and 3.000131. The periods are between 0.6 and 0.7 days. 
Figure~\ref{fig:Halos} shows the four families of Northern and Southern Halo orbits around $L_1$ and $L_2$. The reference frame with axes $\xi$, $\eta$ and $\zeta$ is synodical and centered at Enceladus, and the units are unnormalised. 

\begin{figure}[h!]
\centering
\includegraphics[scale=0.22]{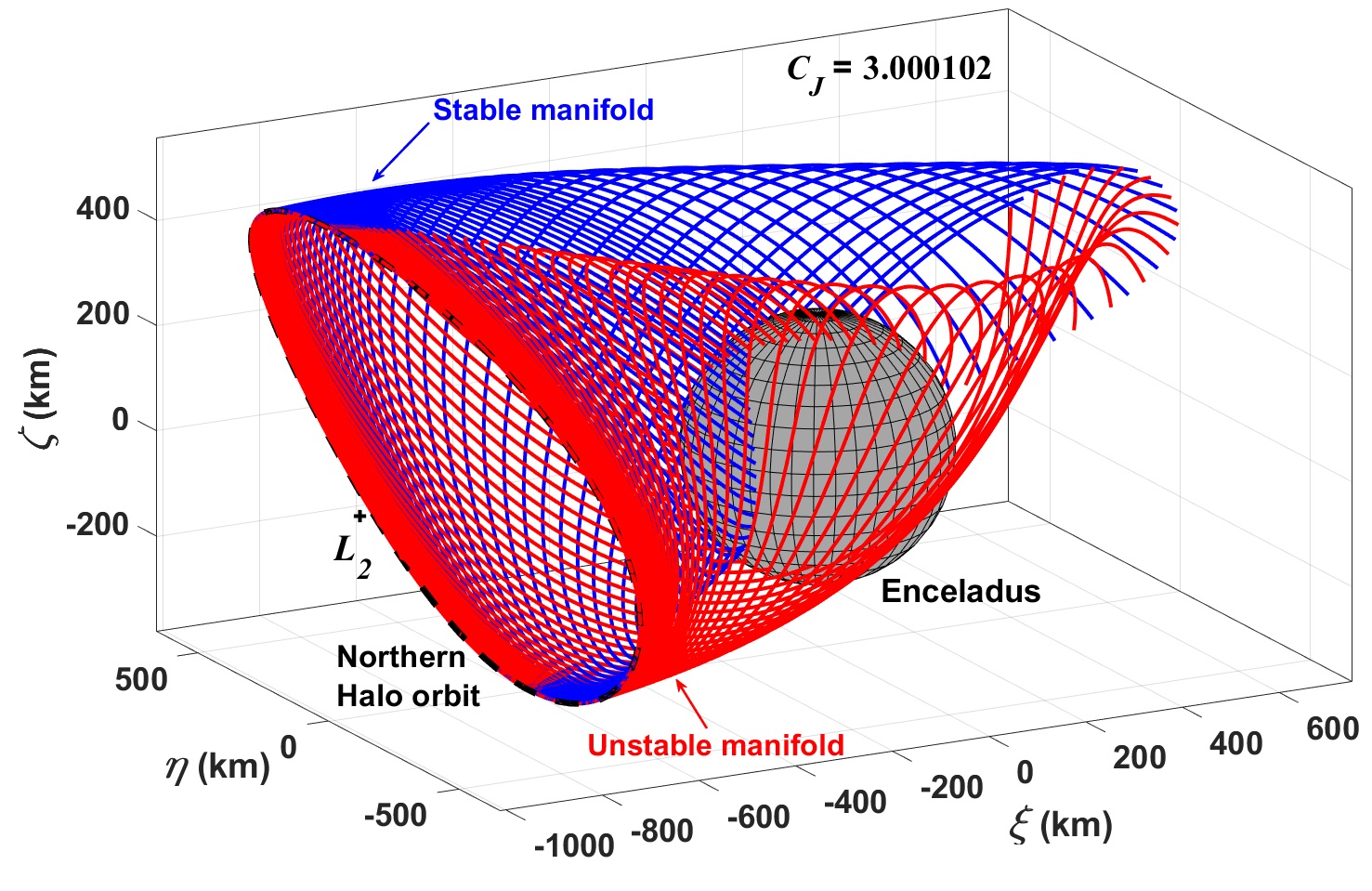} \\
\includegraphics[scale=0.22]{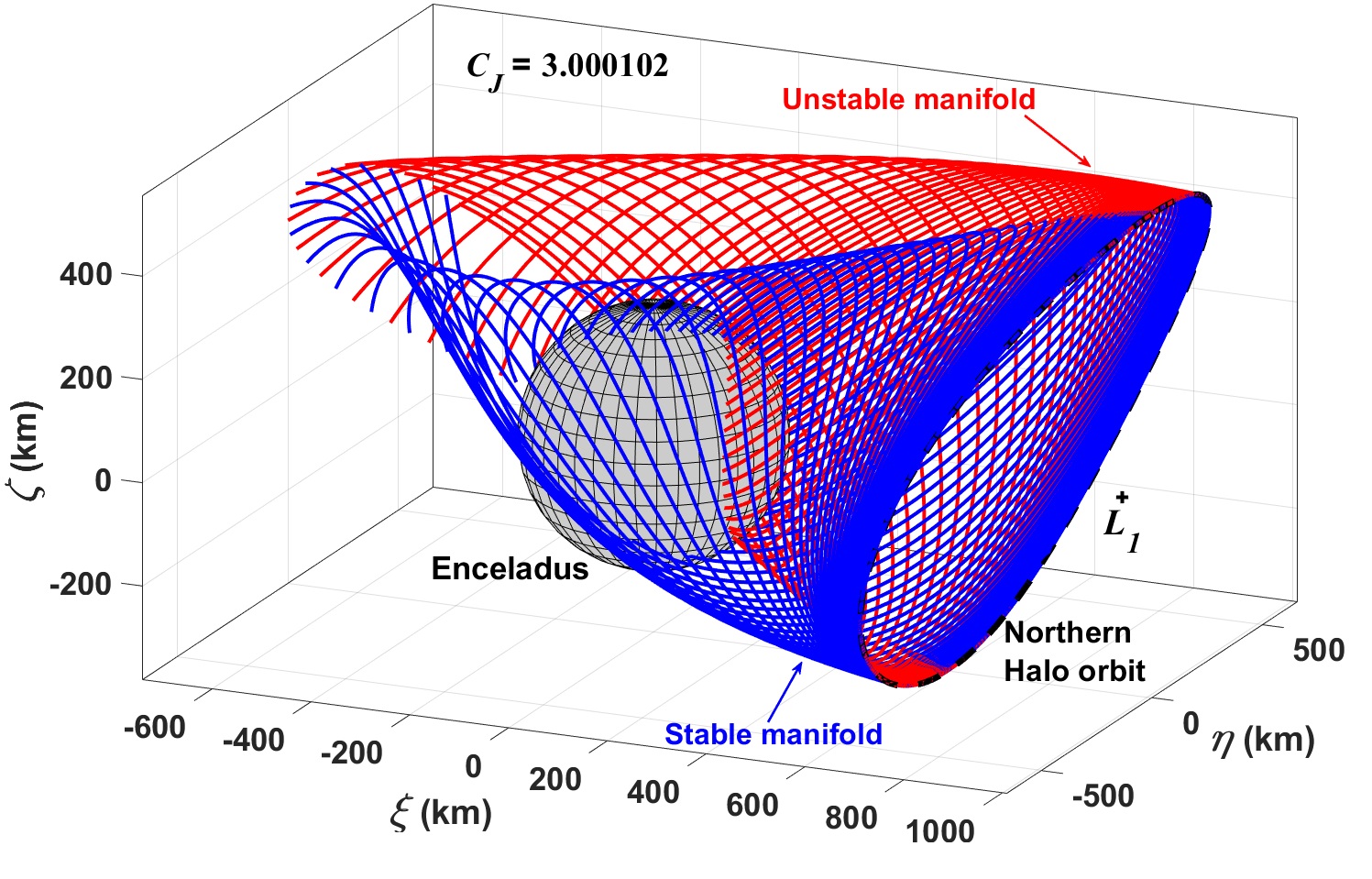}\\
\caption{Branches of the unstable (red) and stable (blue) HIMs of Northern Halo orbits with $C_J$ = 3.000102 around $L_2$ (top) and $L_1$ (bottom) in the Saturn-Enceladus system (Enceladus-centered synodic frame with unnormalised units).}
\label{fig:HIMs}
\end{figure}

The desired branches of the stable and unstable HIMs of the Halo orbits have been approximated by applying a suitable small perturbation in the direction of the stable and unstable eigenvector of the monodromy matrix after appropriate time-transformation through the state transition matrix. Then, these states have been propagated backward and forward in time to globalize the respective trajectories \cite{Parker:1989}. Figure~\ref{fig:HIMs} shows the stable (in blue) and unstable (in red) HIMs of Northern Halo orbits around $L_1$ and $L_2$ with $C_J$ = 3.000102. It has been observed that many trajectories impact Enceladus soon after leaving the vicinity of the Halo orbit as a consequence of the large size of the moon relative to its Hill sphere (respective radii of 252 and 949 km), and this complicates the design of transfers between Halo orbits in the system (see also \cite{Davis:2018}).     
In this exploration, the search for s-heteroclinics has been conducted between orbits with the same Jacobi constant. This facilitates the identification of very low-cost transfers by the method explained below.

\section{Computation of s-heteroclinics between Halo orbits}
\label{sec:heteroclinics}
Intersections between HIMs of different stability character and belonging to different Halo orbits constitute the mechanism here explored to move within the Hill sphere of Enceladus and perform close approaches to the moon.    
In the spatial CR3BP, six state variables characterize a trajectory in phase space. Hence, the above intersection is not easy to determine. Poincar\'e sections are a way of reducing the dimension of a trajectory, thus facilitating its visualization \cite{Koon:2000, Koon:2011}. In the case at hand, the intersections with a plane in configuration space reduces the dimensionality to five. Adding the Jacobi constant as a relationship among variables removes one more dimension.  Hence, eventually the cuts of the HIMs with the plane must be analysed in four dimensions. Different representations have been proposed to visualize four state variables. For example, Haapala \& Howell \cite{Haapala:2013, Haapala:2014} adopted a single segment to represent simultaneously four states: two states are indicated by the coordinates of the segment base-point, and two additional coordinates are represented by the length. Geisel \cite{Geisel:2013} represented $y$, $z$, $\dot{y}$ in a three-dimensional visual environment in which $\dot{z}$ is displayed using color. Paskowitz \& Scheeres \cite{Paskowitz:2006a} and Davis et al. \cite{Davis:2018} chose spherical coordinates to represent the states at the closest approach to the primary (periapsis map).   

\begin{figure}[h!]
\centering
\includegraphics[scale = 0.25]{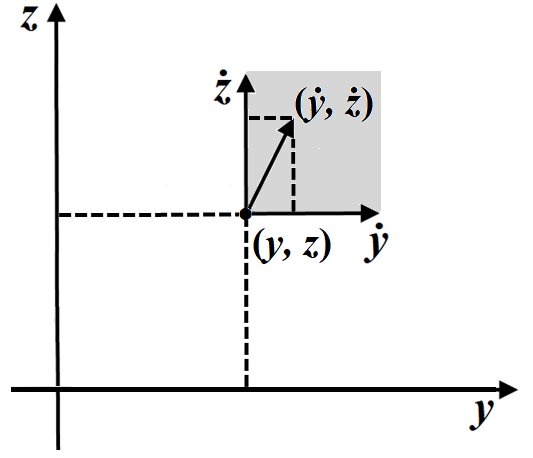}
\caption{Vector representation of the four state variables $y$, $z$, $\dot{y}$, $\dot{z}$.}
\label{fig:sigma}
\end{figure}

In this work, s-heteroclinic connections are designed by propagating the HIMs until their first crossing with the plane $\Sigma$ defined by $x=\mu-1$. $\Sigma$  is orthogonal to the $x$-axis through the center of Enceladus. Numerical experiments showed that this choice is the most suitable to obtain a transversal cut of the flow.  Given the symmetries of the problem,  only transfers for which $\dot{x} > 0$ at the first crossing of the HIMs with $\Sigma$ have been considered.  
Then, a planar visualization based on drawing a vector to represent $y$, $z$, $\dot{y}$, $\dot{z}$ has been adopted: the origin of the vector defines the $yz$-position, whereas its length and orientation indicate the corresponding velocity projection (Fig.~\ref{fig:sigma}). The remaining component of the state, namely $\dot{x}$, is determined by the value of $C_J$. A zero-cost s-heteroclinic connection exists between two Halo orbits with the same $C_J$ when two vectors from their HIMs coincide in position, magnitude and direction in the above described Poincar\'e section. In practice, the solution is built by identifying the trajectories whose vectors differ the least. Such differences constitute the position and velocity error of the solution. A requirement for distance and velocity errors to be respectively less than 1 km and 1 m/s for a connection to be accepted has been introduced. Additionally, a safety distance of 20 km from the surface of Enceladus has been imposed. 

\begin{figure}[h!]
\centering
\includegraphics[scale = 0.35]{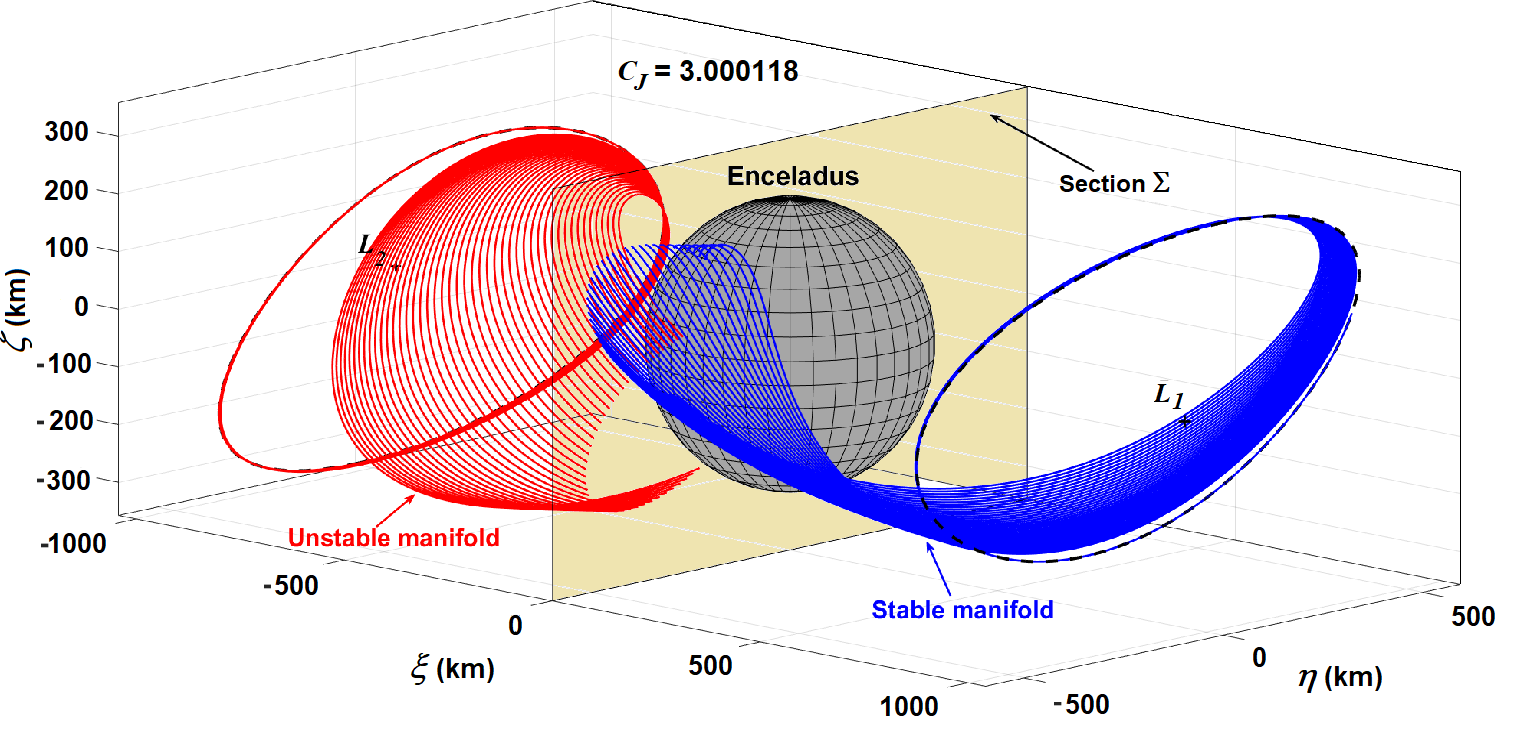}\\
\includegraphics[scale = 0.266]{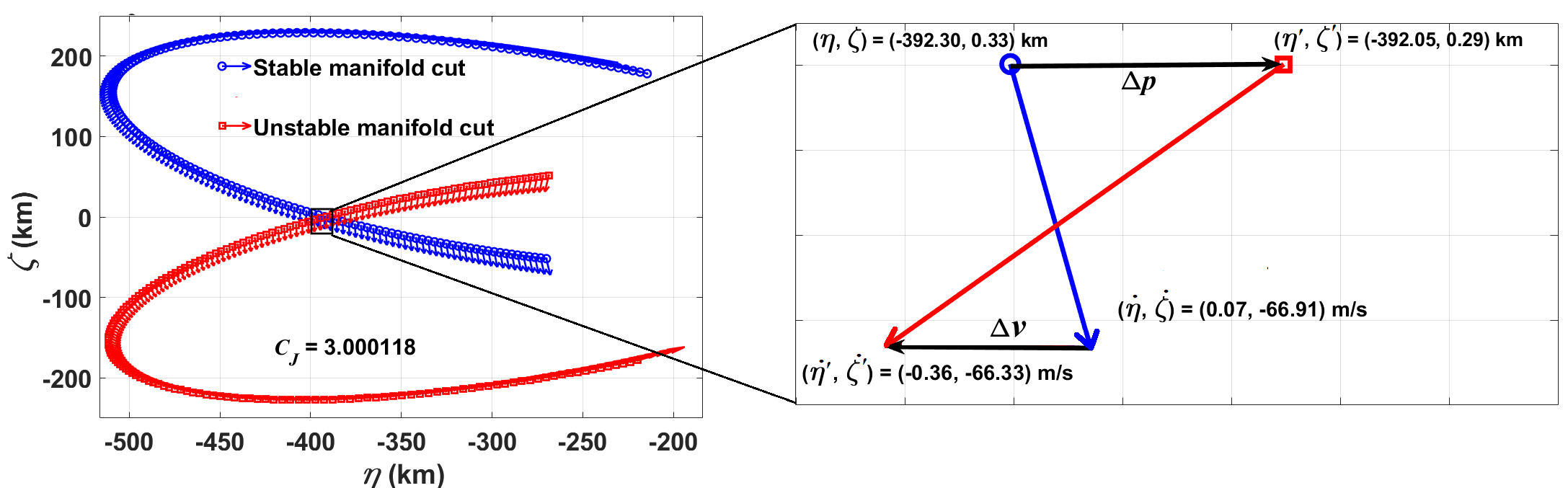}
\caption{Top: non-impacting trajectories of the unstable (red) and stable (blue) HIMs that intersect $\Sigma$ and originate from a Southern Halo orbit around $L_2$ and a Northern Halo orbit around $L_1$, respectively, both with $C_J$ = 3.000118 (Enceladus-centered synodical frame with unnormalised units). Bottom left: vector representation of the intersections with $\Sigma$ (physical units). Bottom right: vectorial representation of the two trajectories that minimize the position and velocity errors.}
\label{fig:example_him}
\end{figure}

\begin{figure}[h!]
\centering
\includegraphics[scale = 0.225]{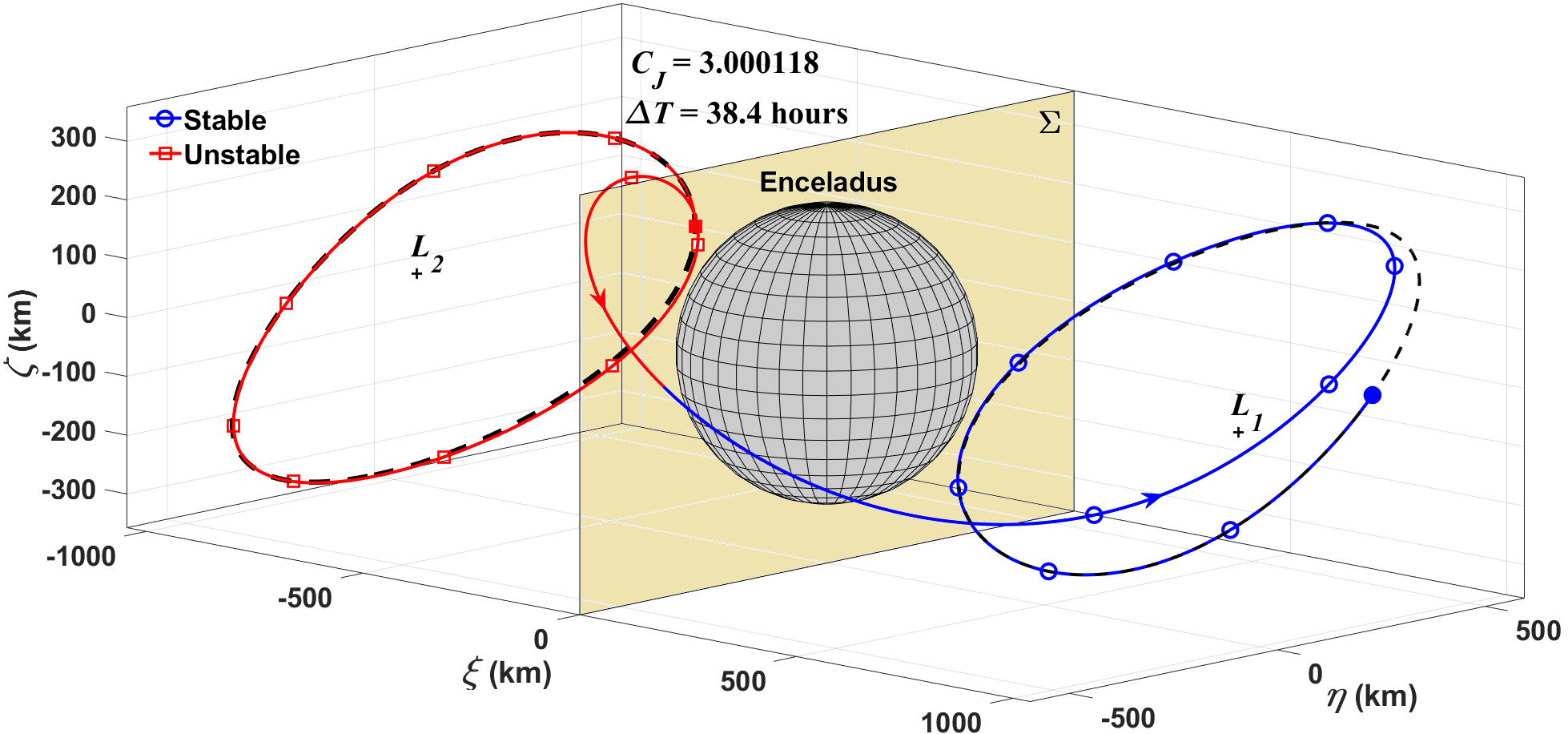}
\caption{3D view of the s-heteroclinic connection from a Southern Halo orbit around $L_2$ to a Northern Halo orbit around $L_1$ with $C_J$ = 3.000118 (Enceladus-centered synodic reference frame with unnormalised units).}
\label{fig:Example_Sol}
\end{figure}

As an example, consider the transfer from a Southern Halo orbit around $L_2$ to a Northern Halo orbit around $L_1$ with $C_J$ = 3.000118. The majority of trajectories (about $65\%$) either impact Enceladus or escape from the Hill sphere. Figure~\ref{fig:example_him} (top) illustrates the trajectories that intersect $\Sigma$. The vector representation of the intersections involving this subset of non-impacting orbits is shown in Fig.~\ref{fig:example_him} (bottom). The blue arrows are associated with the $L_1$ stable manifold, whereas the red arrows represent the $L_2$ unstable manifold. The transfer with the lowest position and velocity error (respectively of 0.26 km and 0.85 m/s) is shown in Fig.~\ref{fig:Example_Sol}. The time of flight $\Delta T$ from Halo to Halo is 38.4 hours.   

Owing to the symmetries of the model, four types of connections between Northern and Southern families need to be explored in order to account for all the possible combinations of departure and arrival orbits. These are: Northern $L_1$ to Northern $L_2$ (type A), Southern $L_2$ to Northern $L_1$ (type B), Northern $L_1$ to Southern $L_1$ (type C) and Southern $L_2$ to Northern $L_2$ (type D). 

The application of the above method to the families of Halo orbits described in Sect.~\ref{sec:model} has allowed to identify four solutions with position and velocity errors below the chosen threshold: two transfers between $L_1$ and $L_2$ with $C_J$ = 3.000118, and two transfers with $C_J$ = 3.000072 between Northern and Southern orbits around the same libration point. 
It is worth noting that the solution of type A of this investigation is in good agreement with the result of Davis et al. \cite{Davis:2018}.

\begin{figure}[h!]
\centering
\includegraphics[scale = 0.125]{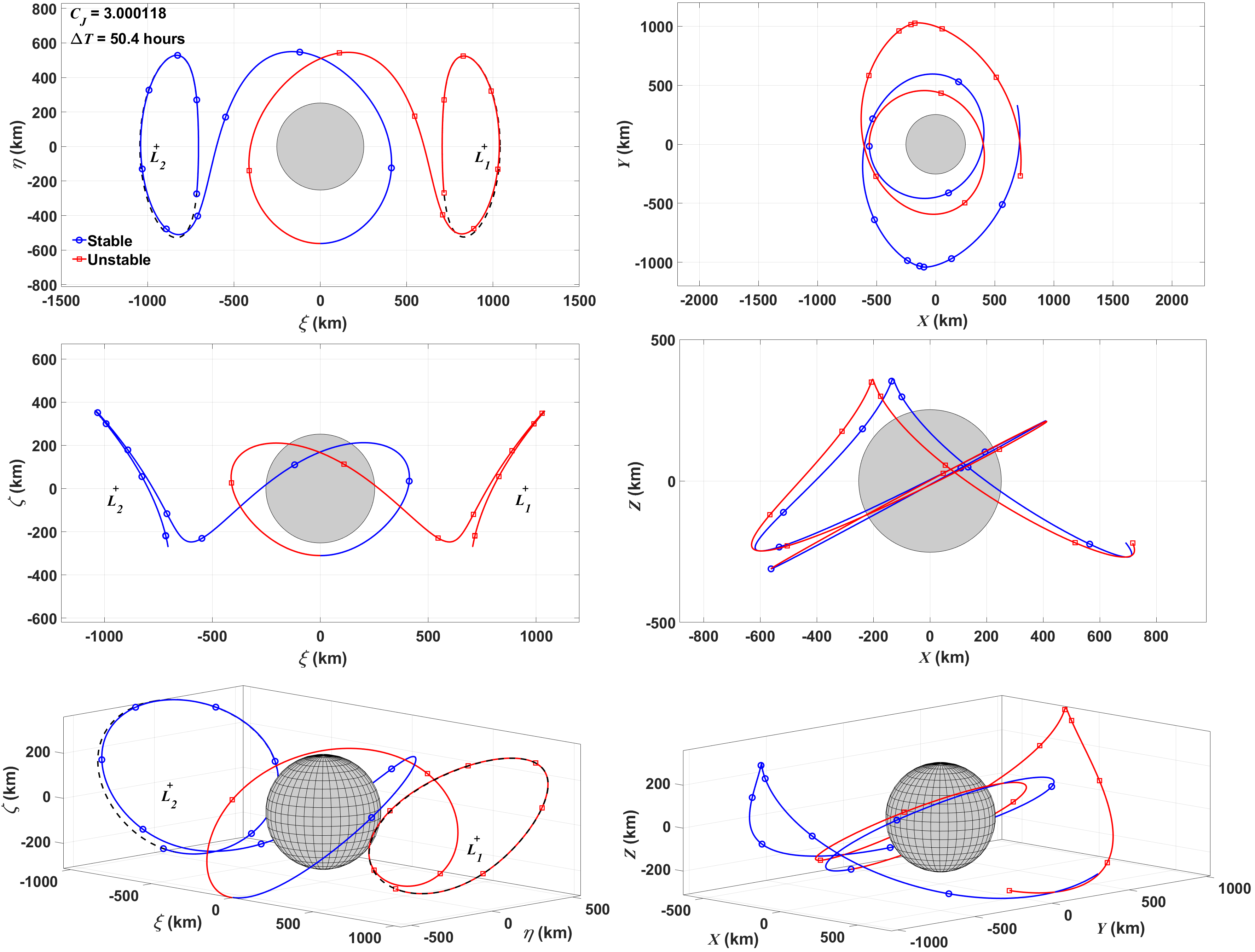}
\caption{S-heteroclinic connection of type A with $C_J$ = 3.000118: Enceladus-centered unnormalised synodical frame (left), Enceladus-centered inertial frame (right), planar projections (top and middle), 3D view (bottom).}
\label{fig:Type_A}
\end{figure}

\begin{figure}[h!]
\centering
\includegraphics[scale = 0.125]{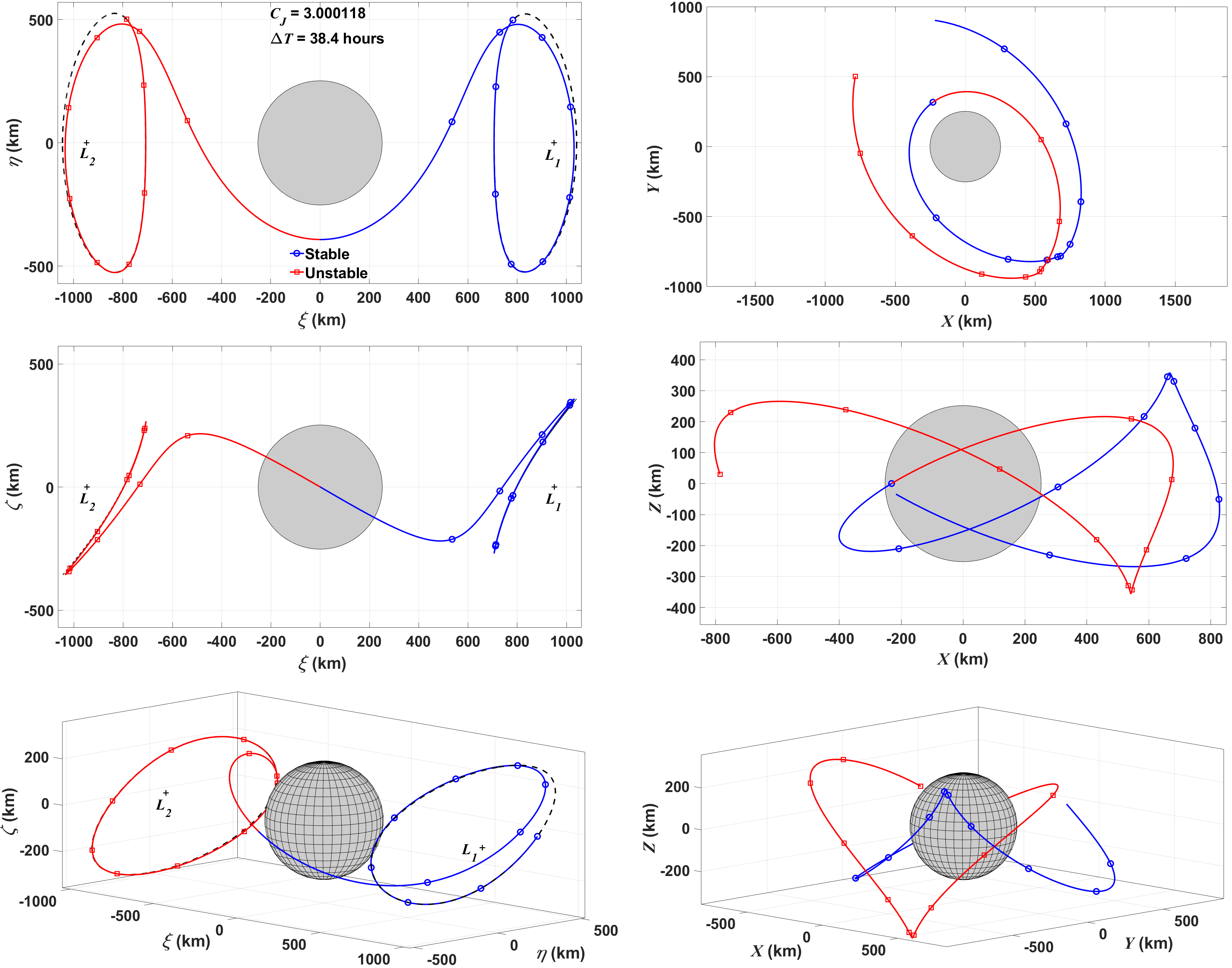}
\caption{S-heteroclinic connection of type B with $C_J$ = 3.000118: Enceladus-centered unnormalised synodical frame (left), Enceladus-centered inertial frame (right), planar projections (top and middle), 3D view (bottom).}
\label{fig:Type_B}
\end{figure}

\begin{figure}[h!]
\centering
\includegraphics[scale = 0.125]{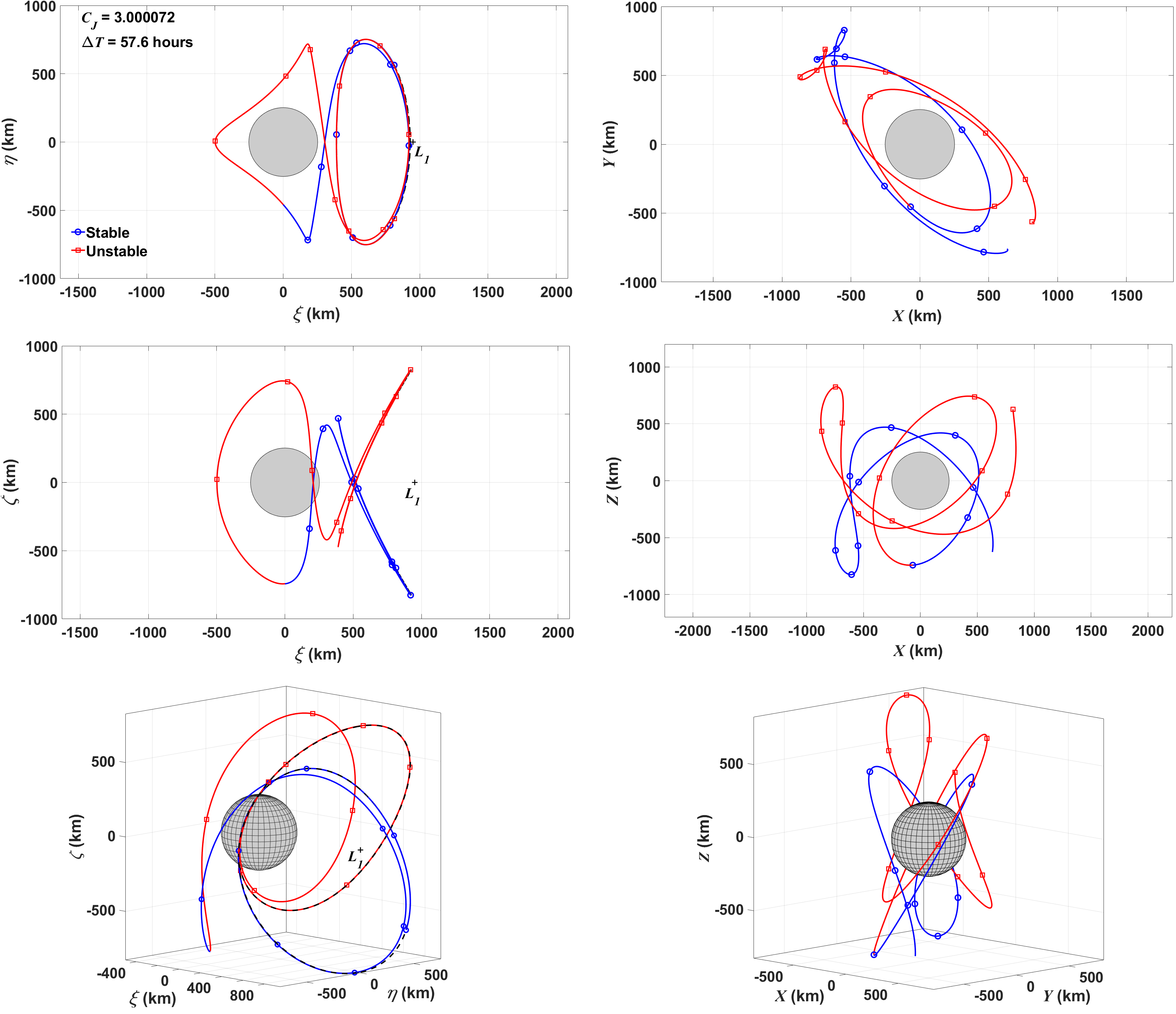}
\caption{S-heteroclinic connection of type C with $C_J$ = 3.000072: Enceladus-centered unnormalised synodical frame (left), Enceladus-centered inertial frame (right), planar projections (top and middle), 3D view (bottom).}
\label{fig:Type_C}
\end{figure}

\begin{figure}[h!]
\centering
\includegraphics[scale = 0.125]{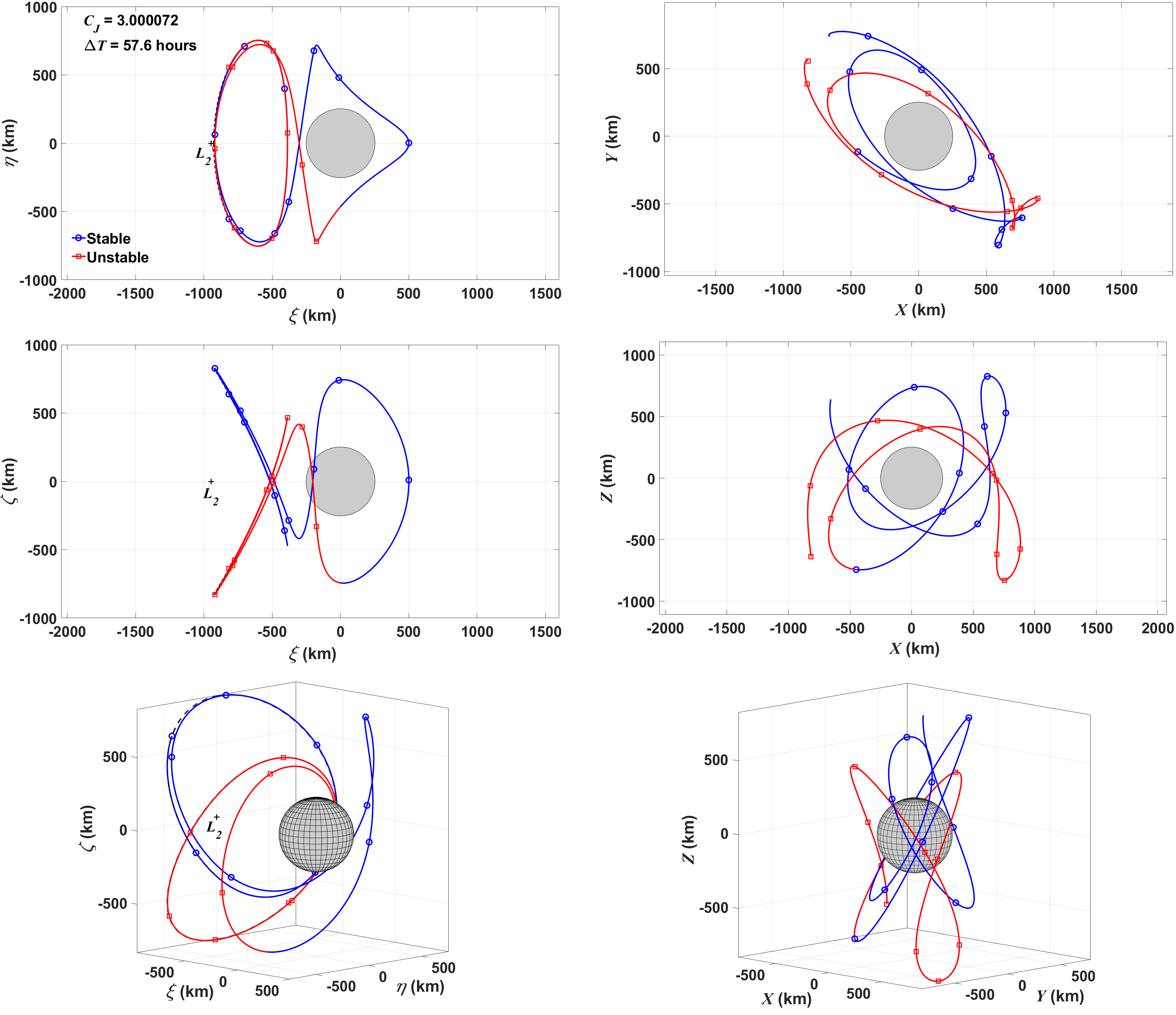}
\caption{S-heteroclinic connection of type D with $C_J$ = 3.000072: Enceladus-centered unnormalised synodical frame (left), Enceladus-centered inertial frame (right), planar projections (top and middle), 3D view (bottom).}
\label{fig:Type_D}
\end{figure}

The time of flight along the solutions ranges from 38.4 to 57.6 days. Table~\ref{tab:conn} summarizes the basic features of these s-heteroclinics, i.e., connection type, Jacobi constant, time of flight, and distance and velocity errors.
An Enceladus-centered reference frame with inertial axes $X$, $Y$, $Z$ has been introduced. $X$, $Y$, $Z$ are assumed to be parallel to $x$, $y$, $z$ at the beginning of a transfer ($t$=0). 
Figures~\ref{fig:Type_A} to \ref{fig:Type_D} show the four trajectories. Each figure contains six plots: two planar projections and the 3D trajectory in the synodical (left) and in the inertial (right) reference frames centered at Enceladus. Open circles and squares have been drawn at constant intervals of time along the stable and unstable portions of the trajectory, respectively. 
     
\begin{figure}[h!]
\centering
\includegraphics[scale = 0.23]{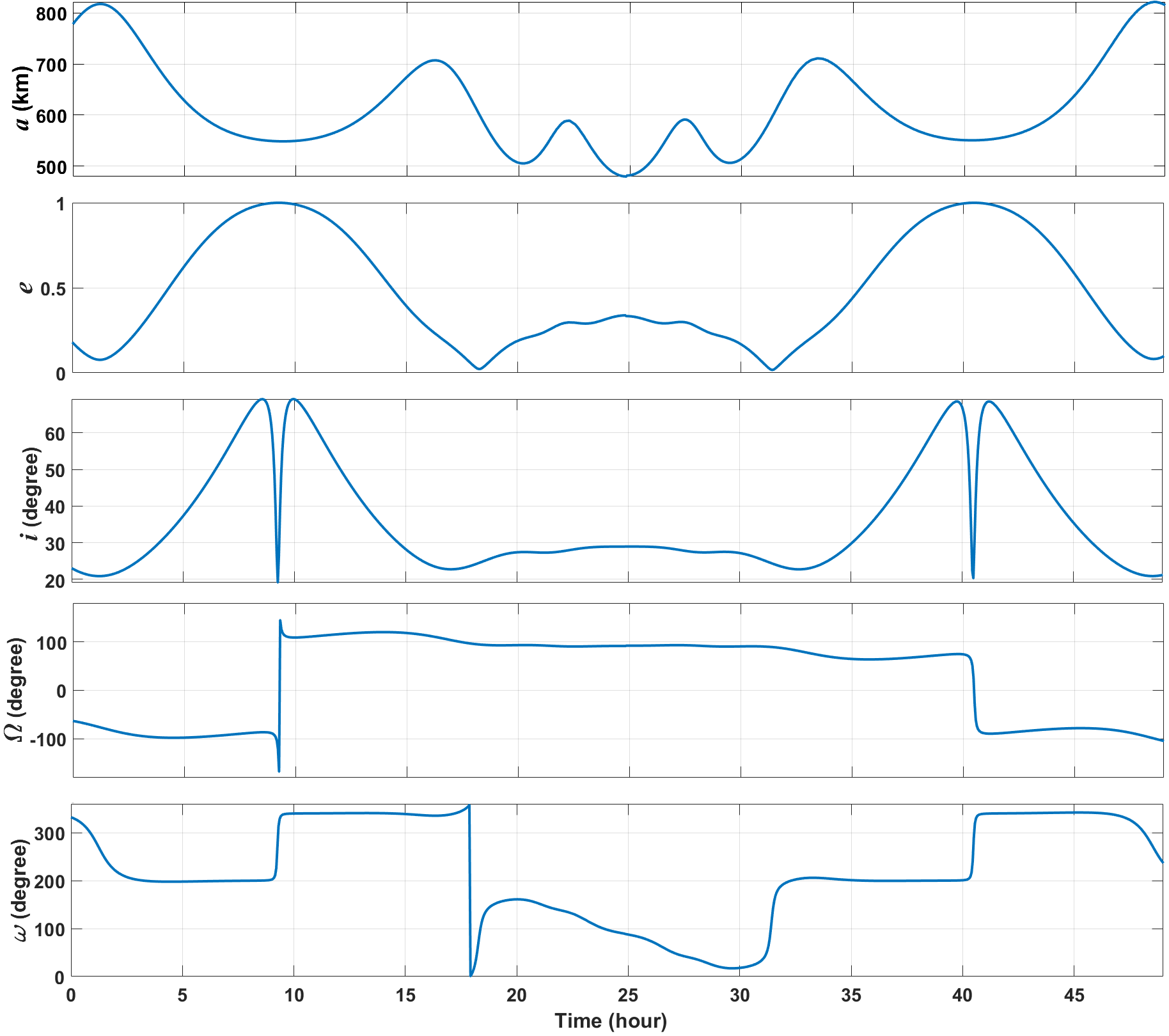}
\caption{Evolution of the osculating orbital elements for the trajectory of type A shown in Fig.~\ref{fig:Type_A} (Enceladus-centered reference frame with inertial axes).}
\label{fig:OrbEl_Type_A}
\end{figure}

\begin{figure}[h!]
\centering
\includegraphics[scale = 0.23]{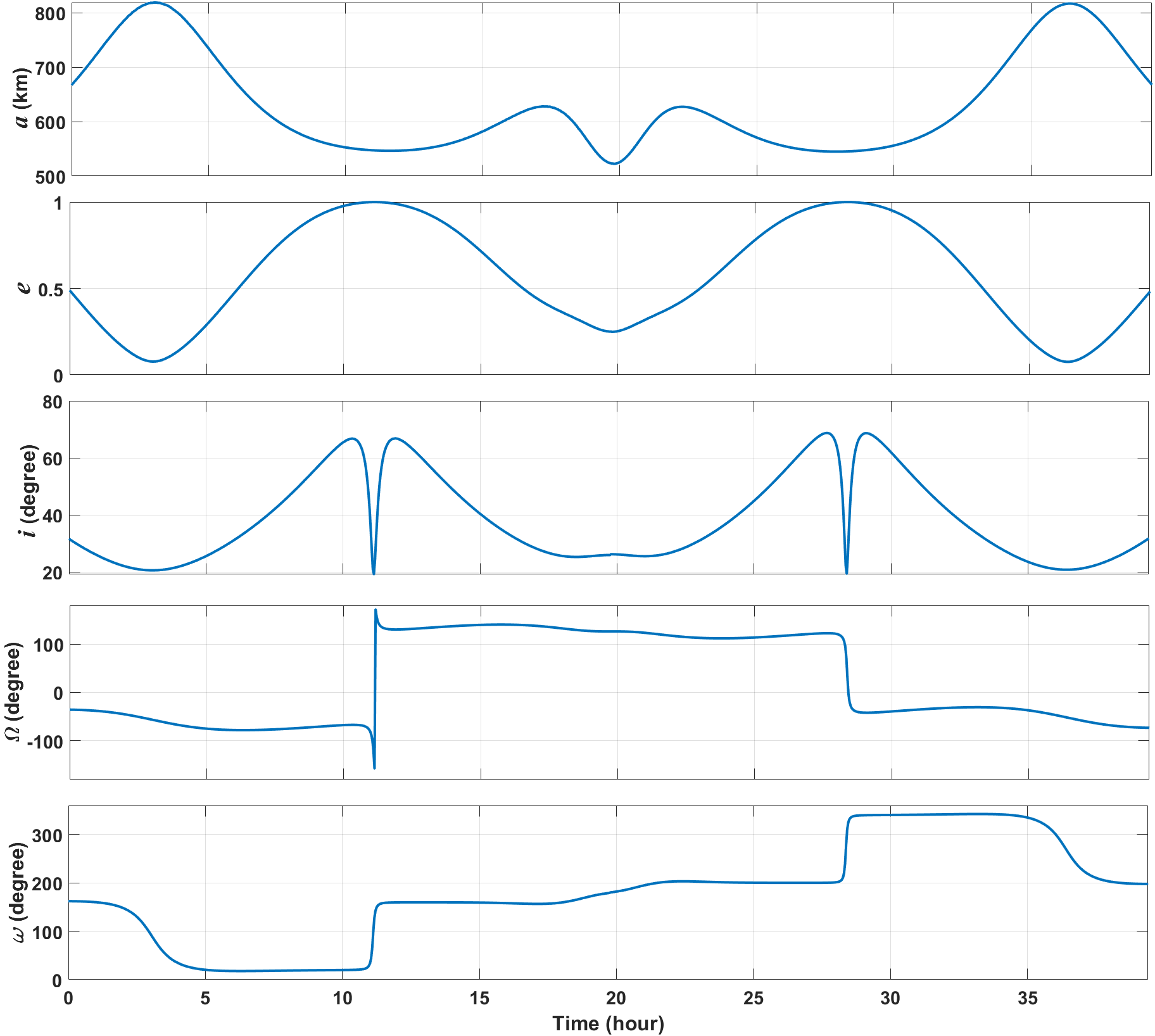}
\caption{Evolution of the osculating orbital elements for the trajectory of type B shown in Fig.~\ref{fig:Type_B} (Enceladus-centered reference frame with inertial axes).}
\label{fig:OrbEl_Type_B}
\end{figure}

\begin{figure}[h!]
\centering
\includegraphics[scale = 0.23]{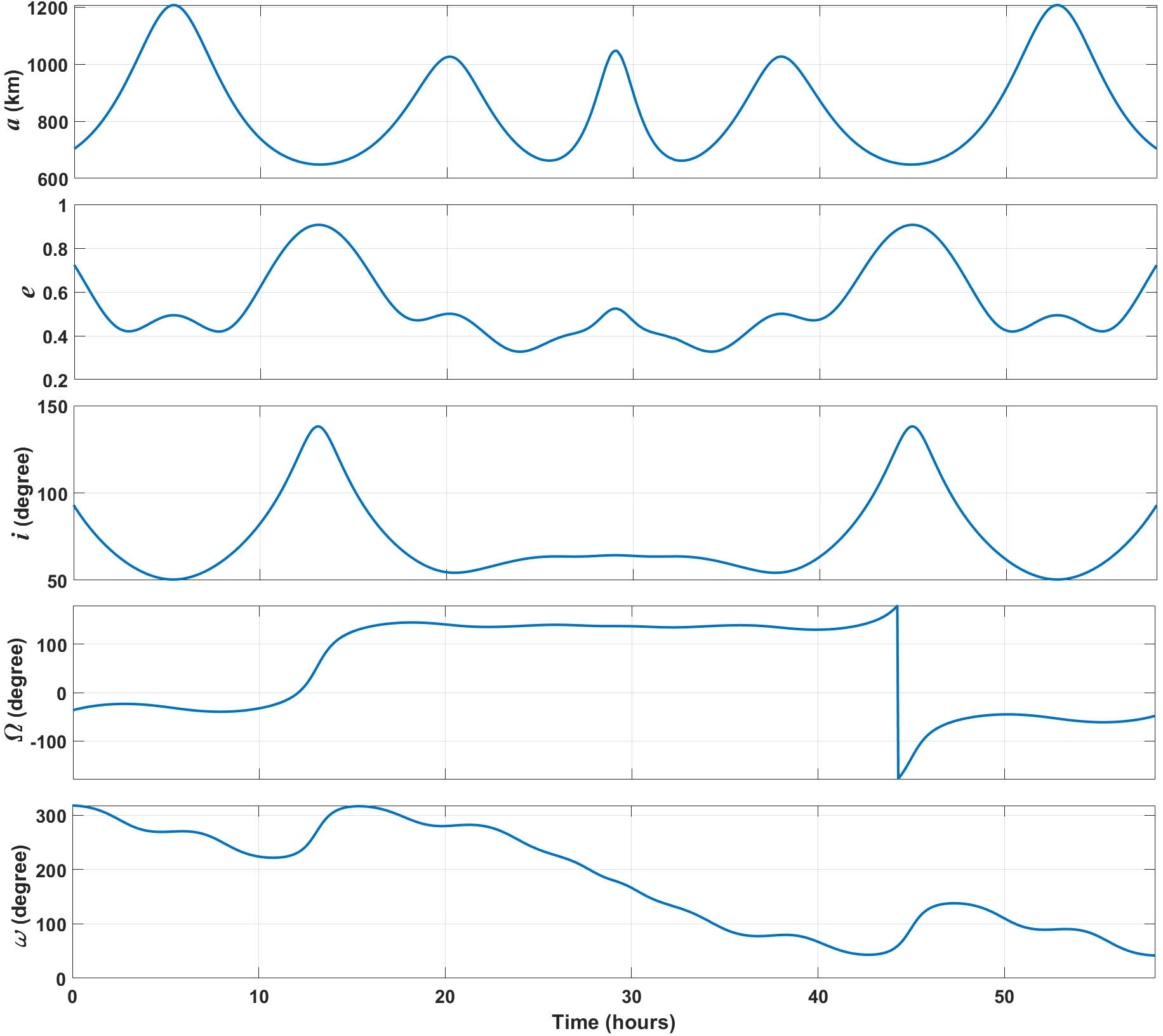}
\caption{Evolution of the osculating orbital elements for the trajectories  of type C and D shown in Figs.~\ref{fig:Type_C} and \ref{fig:Type_D} (Enceladus-centered reference frame with inertial axes).}
\label{fig:OrbEl_Type_C}
\end{figure}

The representation of the trajectories of Figs.~\ref{fig:Type_A} to \ref{fig:Type_D} in the ($X$, $Y$, $Z$) frame allows to determine the evolution of their osculating orbital elements, i.e., semimajor axis $a$, eccentricity $e$, inclination $i$, argument of periapsis $\omega$, longitude of the ascending node $\Omega$ (Figs.~\ref{fig:OrbEl_Type_A} to \ref{fig:OrbEl_Type_C}). All elements vary considerably as a result of the perturbing effect of Saturn's gravity. It is interesting to note the wide range of values covered by inclination and eccentricity, the latter reaching as high as 1, i.e., the escape condition.  
\begin{table}[h!]
\begin{center}
\caption{Features of the s-heteroclinic connections between Halo orbits in the Saturn-Enceladus CR3BP: connection type, Jacobi constant, time of flight $\Delta T$ from Halo to Halo, distance error $\Delta p$ and velocity error $\Delta v$ at $\Sigma$.
\vspace{5mm}} 
\label{tab:conn}
\begin{tabular}{ccccc} 
\hline\noalign{\smallskip} 
Connection type & $C_J$   &  $\Delta T$  & $\Delta p$    & $\Delta v$  \\
               &          & (hour)       &  (km)         & (m/s)       \\																		
\noalign{\smallskip}\hline\noalign{\smallskip}
A: Northern $L_1$ to Northern $L_2$ & 3.000118   & 50.4                &  0.41         &  0.52\\
B: Southern $L_2$ to Northern $L_1$ & 3.000118   & 38.4                &  0.26         &  0.85\\
C: Northern $L_1$ to Southern $L_1$ & 3.000072   & 57.6                &  0.75         &  0.23\\
D: Southern $L_2$ to Northern $L_2$ & 3.000072   & 57.6                &  0.23         &  0.17\\
\noalign{\smallskip}\hline
\end{tabular}
\end{center}
\end{table}


\section{Observational performance}
\label{sec:observ}
\begin{figure}[h!]
\centering
\includegraphics[scale = 0.21]{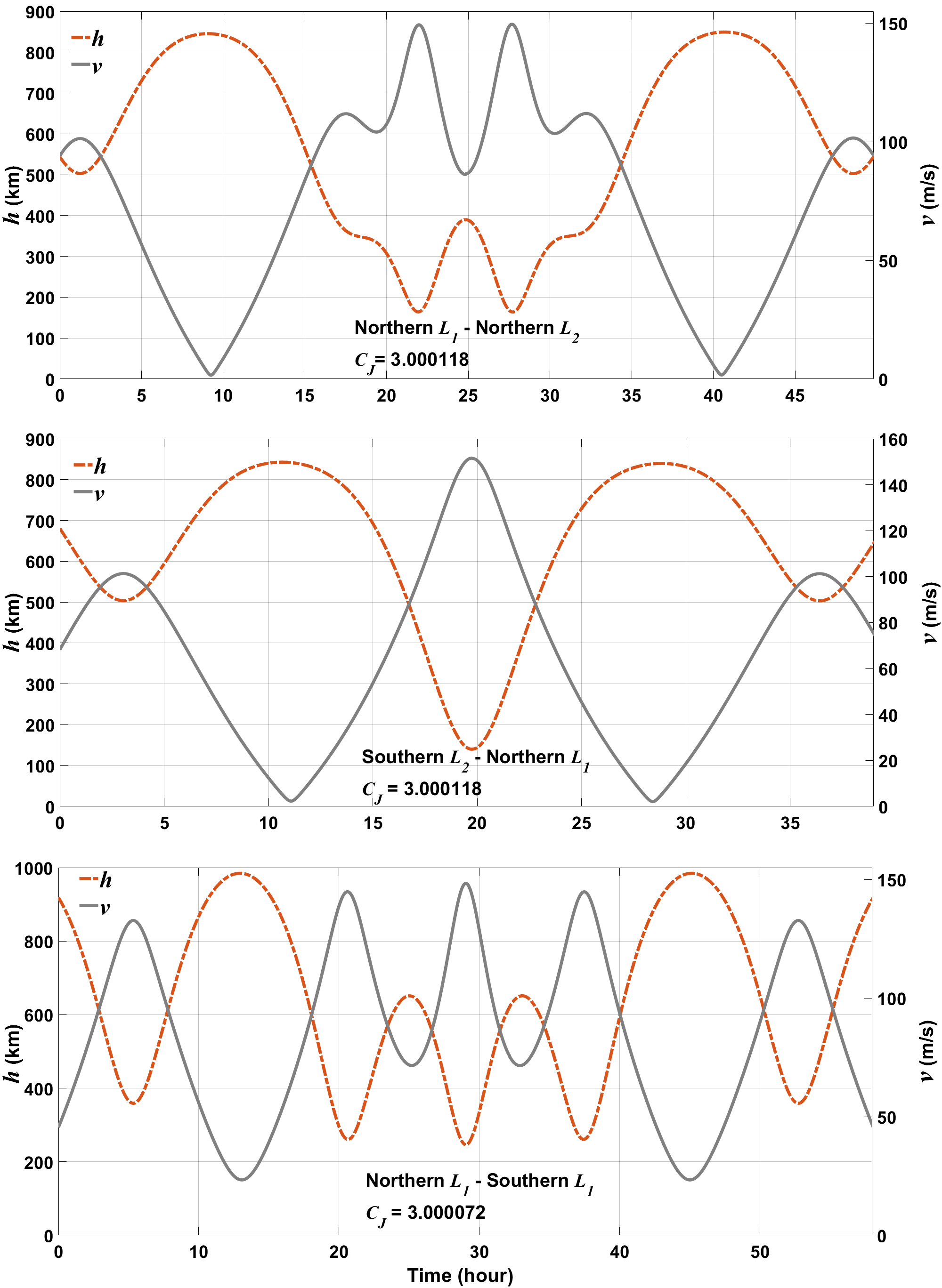}
\caption{Time history of the altitude $h$ above Enceladus and the inertial velocity $v$ over the solutions of Figs.~\ref{fig:Type_A} (top), \ref{fig:Type_B} (middle), and \ref{fig:Type_C}-\ref{fig:Type_D} (bottom).}
\label{fig:h_v}
\end{figure}

Figure~\ref{fig:h_v} shows the time history of the altitude $h$ and the magnitude $v$ of the inertial velocity of the S/C over the s-heteroclinic connections of Figs.~\ref{fig:Type_A} to \ref{fig:Type_D}. The minimum altitude above the lunar surface is 150 km for the solutions with $C_J$ = 3.000118, and approximately 300 km for those with $C_J$ = 3.000072, whereas the maximum altitude is 850 km and 1000 km, respectively. The velocity is always lower than $150$ m/s.                
\begin{figure}[h!]
\centering
\includegraphics[scale = 0.45]{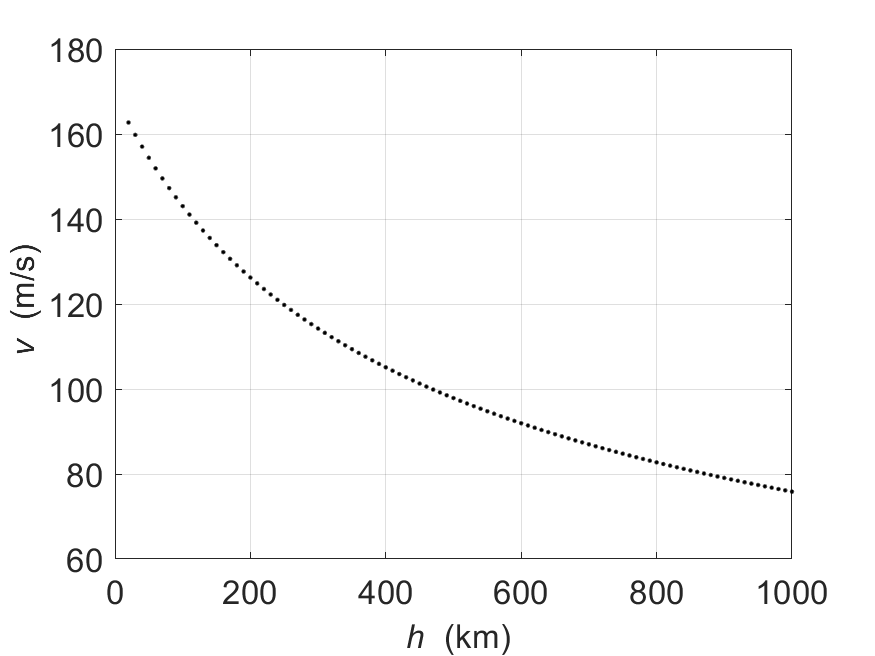}
\caption{Magnitude of the velocity for Keplerian circular orbits around Enceladus as a function of the altitude above the surface.}
\label{fig:v_h_Kepler}
\end{figure}
For the sake of comparison, Fig.~\ref{fig:v_h_Kepler} shows the magnitude of the circular velocity on Keplerian orbits in the same range of altitudes as the solutions here discussed. 

\begin{figure}[h!]
\centering
\includegraphics[scale = 0.495]{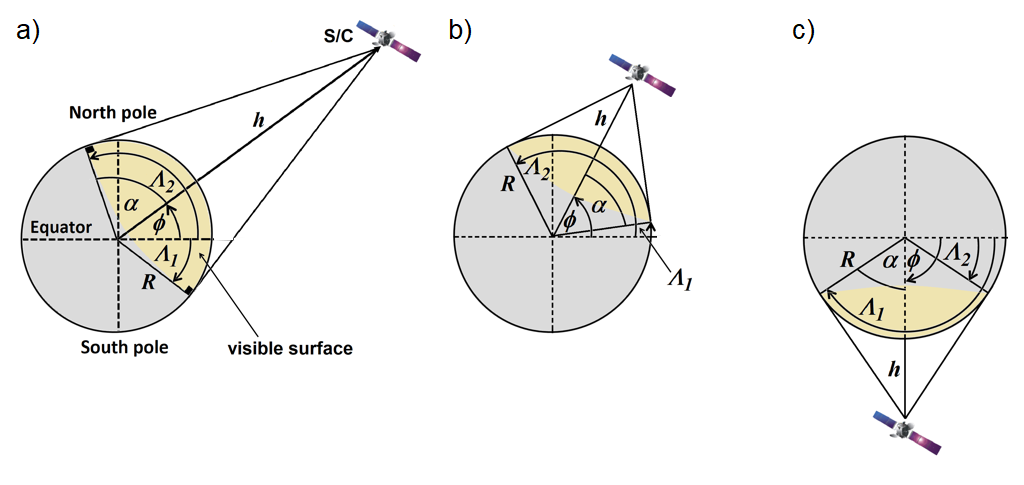}
\caption{Definition of the instantaneous coverage parameters $\phi$, $\alpha$, $h$, $\Lambda_1$ and $\Lambda_2$ and their value for three different S/C positions.}
\label{fig:coverage}
\end{figure}

The instantaneous coverage of the surface of the moon can be quantified by the parameters $\Lambda_1$ and $\Lambda_2$ shown in Fig.~\ref{fig:coverage}a. They represent the limits of the central angle of coverage of amplitude $2\alpha$ and are measured positively northwards from the equator. The angle $\alpha$ depends on the equatorial radius $R$ and the altitude $h$ of the S/C through
\begin{equation}
\alpha = \cos^{-1}\left(\frac{R}{R+h}\right).
\end{equation}
If $\phi$ denotes the latitude of the S/C, then $\Lambda_1$ and $\Lambda_2$ are defined as
\begin{eqnarray}
\Lambda_1 & = & \phi - \alpha, \\
\Lambda_2 & = & \phi + \alpha.
\label{eq:3}
\end{eqnarray}
For example, when $\phi$ = $40^{\circ}$ and $h$ = 500 km, $\alpha$ = $70.4^{\circ}$, $\Lambda_1$ = $-30.4^{\circ}$, $\Lambda_2$ = $110.4^{\circ}$ and the instantaneous coverage extends from below the equator till beyond the north pole (Fig.~\ref{fig:coverage}a). For $\phi$ = $65^{\circ}$ and $h$ = 200 km, $\alpha$ = $56.1^{\circ}$, $\Lambda_1$ = $8.9^{\circ}$, $\Lambda_2$ = $121.1^{\circ}$ and the S/C can see part of the northern hemisphere including the pole but does not see the equator (Fig.~\ref{fig:coverage}b). Eventually, for a polar view with $\phi$ = $-90^{\circ}$ and $h$ = 200 km, $\alpha$ = $56.1^{\circ}$, $\Lambda_1$ = $-146.1^{\circ}$, $\Lambda_2$ = $-33.9^{\circ}$ and the visible area is centered around the south pole (Fig.~\ref{fig:coverage}c). Therefore, when the interval [$\Lambda_1$,$\Lambda_2$] includes either $-90^{\circ}$ or $+90^{\circ}$, the S/C has access to the corresponding pole.

\begin{figure}[h!]
\centering
\includegraphics[scale = 0.23]{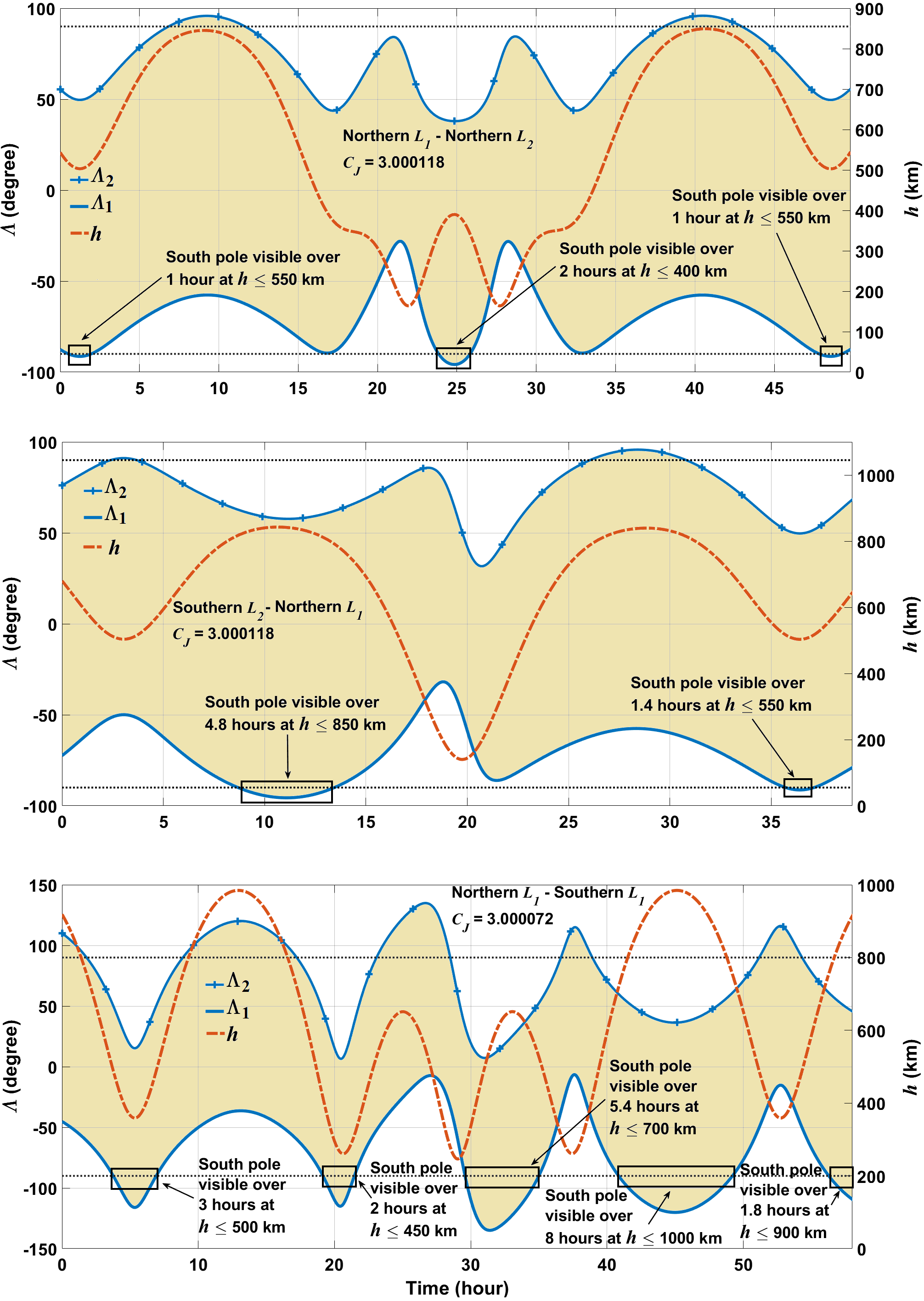}
\caption{Time history of the instantaneous coverage parameters $\Lambda_1$, $\Lambda_2$ and the altitude $h$ as functions of time over the transfers of Figs.~\ref{fig:Type_A} (top), \ref{fig:Type_B} (middle), and \ref{fig:Type_C}-\ref{fig:Type_D} (bottom).}
\label{fig:lambda_hist}
\end{figure}

The time history of $\Lambda_1$ and $\Lambda_2$ for the transfers of Figs.~\ref{fig:Type_A} to \ref{fig:Type_D} is visualised in 
Fig.~\ref{fig:lambda_hist} (this figure contains three plots because the results for Figs.~\ref{fig:Type_C} and \ref{fig:Type_D} are identical). The significant out-of-plane motion of these trajectories allows observation of both polar regions. The yellow areas in Fig.~\ref{fig:lambda_hist} represent the instantaneous amplitude of coverage as a function of time, limited by the curves of $\Lambda_1$ and $\Lambda_2$. The plots also show the altitude of the S/C, whereas the horizontal dashed lines indicate the two poles. The total duration of the observation windows for the south pole amounts to approximately 4 hours per transfer when $C_J = 3.000118$ (type A), just above 6 hours when $C_J = 3.000118$ (type B) and $21$ hours when $C_J = 3.000072$ (types C and D). 

\begin{figure}[h!]
\centering
\includegraphics[scale = 0.22]{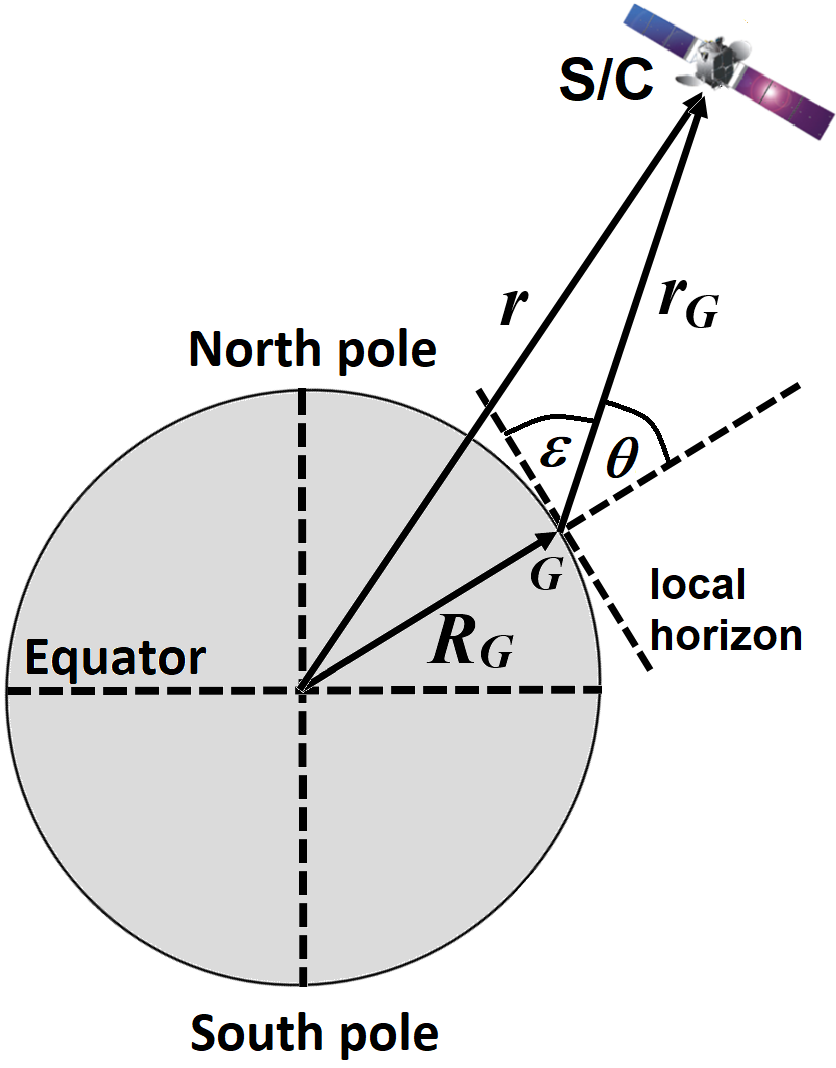}
\caption{Definition of local horizon for a point (G) on the surface of Enceladus, and the corresponding elevation $\epsilon$ and co-elevation $\theta$ of the S/C.}
\label{fig:horizon}
\end{figure}

\begin{figure}[h!]
	\centering
	\includegraphics[scale = 0.20]{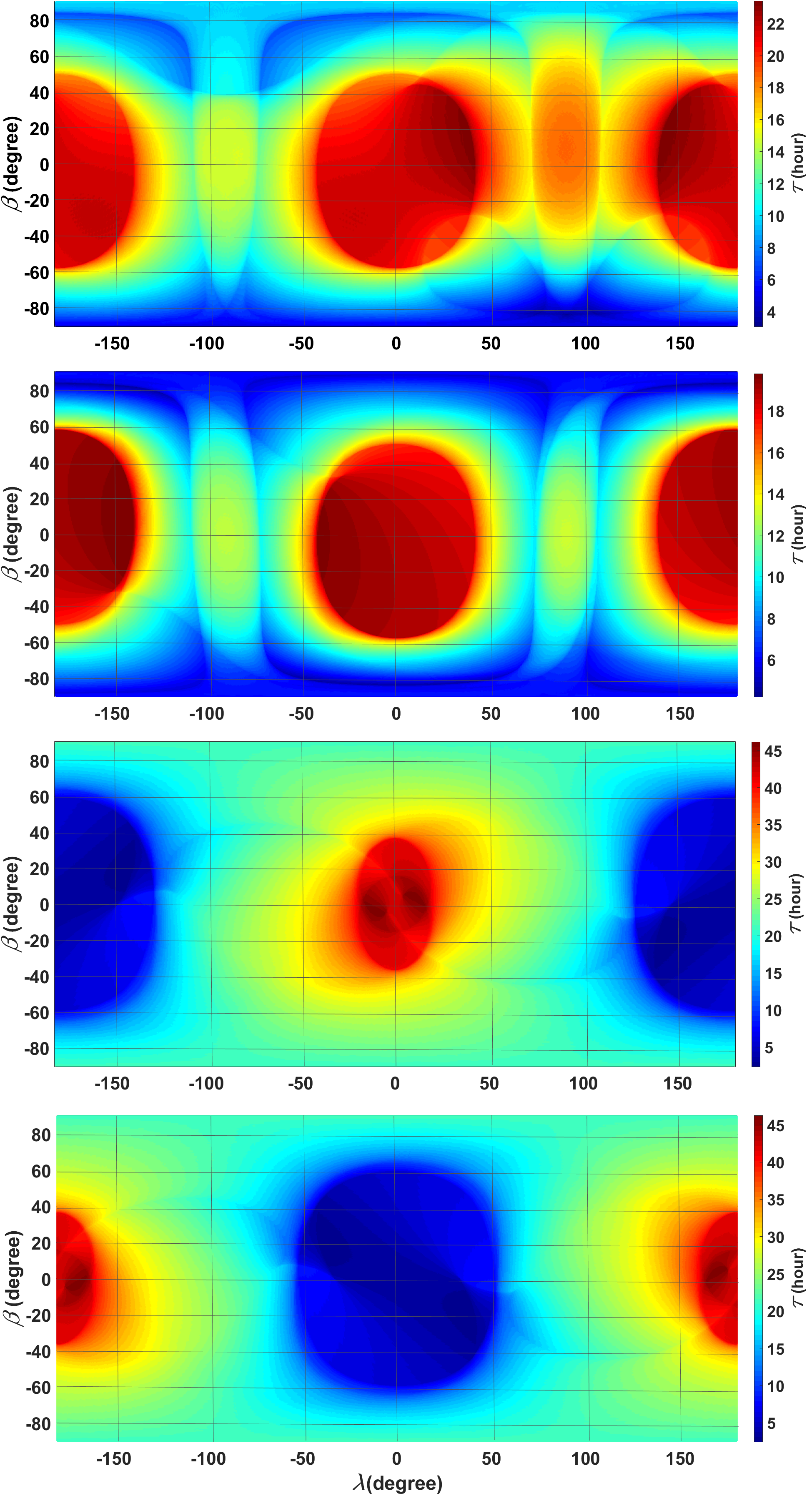}
	\caption{Geographical maps of total time of overflight for the connections of Figs.~\ref{fig:Type_A} to \ref{fig:Type_D}, respectively from top to bottom.}
	\label{fig:tau}
\end{figure}

Additionally, we have computed the total time of overflight $\tau$ (defined as the total access time of a specific surface point) for the entire surface of Enceladus on each transfer. This parameter depends on the visibility of a given point from the S/C and this, in turn, is expressed by the condition $\epsilon \ge 0$, where  $\epsilon$ is the elevation of the S/C on the local horizon (see Fig.~\ref{fig:horizon}). By discretising $\Delta T$ in $N$ intervals of duration $\delta t$ and assigning to each an elementary time of overflight $\delta \tau_i$ ($i$ = 1,2,...,$N$),
\begin{equation}
\label{eq:3}
\delta\tau_i =
\begin{cases}
\delta t & \text{if } \epsilon_i \geq 0
\\
0 & \text{otherwise,} 
\end{cases}
\end{equation}
yields the total time of overflight $\tau$ at the given location as
\begin{equation}
\label{eq:4}
\tau = \sum_{i=1}^{N}\delta\tau_i.
\end{equation}
The computation of $\tau$ has been carried out in the Enceladus-centered synodical frame, in this way taking into account the effect of the spin of the moon which is synchronous with its orbital motion. Thus, point G (Fig.~\ref{fig:horizon}) is stationary and its components are given by
\begin{equation}
\label{eq:5}
{\bf R}_G = R \left(\cos{\lambda}\cos{\beta}, \sin{\lambda}\cos{\beta}, \sin{\beta}\right)^T,
\end{equation}
with $\lambda$ and $\beta$ the geographical longitude and latitude of $G$. 
Then, the co-elevation $\theta$ of the S/C from G is obtained through
\begin{equation}
\label{eq:7}
\cos \theta = {\bf R}_G \cdot {\bf r}_G = {\bf R}_G \cdot \left({\bf r} - {\bf R}_G \right),
\end{equation}
with ${\bf r}$ = $\left(\xi,\eta,\zeta\right)^T$ the Enceladus-centered synodical position vector of the S/C at the given time. Eventually,
\begin{equation}
\label{eq:6}
\epsilon = 90^{\circ} - \theta.
\end{equation}
The geographical maps of $\tau$ reported in Fig.~\ref{fig:tau} have been obtained by discretising the surface of Enceladus at intervals of 0.01 radians in longitude and latitude. 

\begin{figure}[h!]
	\centering
	\includegraphics[scale = 0.22]{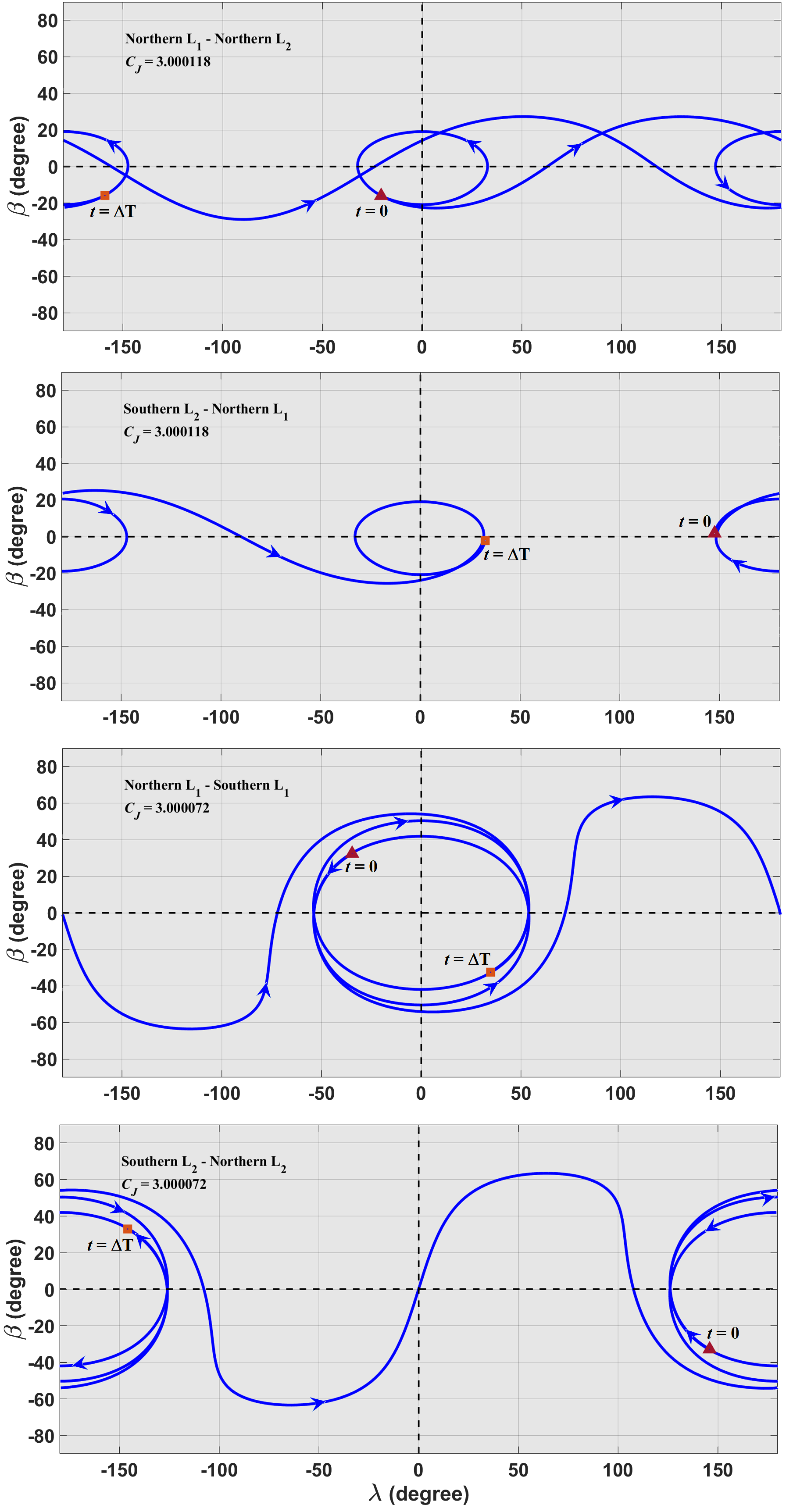}
	\caption{Ground tracks for the trajectories of Figs.~\ref{fig:Type_A} to \ref{fig:Type_D}, respectively from top to bottom.}
	\label{fig:GT}
\end{figure}

Eventually, we have computed the ground track of the S/C, i.e., the projection of the S/C's orbit onto the surface of Enceladus.
Figure~\ref{fig:GT} shows the result for the four solutions of this study.

\section{Discussion and conclusions}
\label{sec:concl}
The discovery of geyser-like jets from the south pole of Enceladus makes this moon remarkably interesting from the scientific point of view. This fact is boosting plans for the robotic exploration of this realm. The need for suitable science orbits cannot be satisfied by conventional moon-S/C two-body solutions, given the strong instabilities caused in this case by the gravity of the planet. Following the path traced by other contributions in the framework of the CR3BP and Hill's model, this work proposes the use of periodic solutions around libration points, in particular Halo orbits, to generate
low-energy trajectories for scientific purposes. Such trajectories are s-heteroclinic connections between Northern and Southern Halo orbits around $L_1$ and $L_2$ of the CR3BP formed by Saturn, Enceladus and the S/C. The search for these connections has been carried out in a systematic way by varying the energy of the progenitor Halo orbits. A planar, vectorial visualization of position and velocity components has facilitated the identification of intersections between HIMs originating from two orbits at a suitable Poincar\'e section. 
At the adopted energy discretisation (100 Halo orbits in each family covering a Jacobi constant range between 3.000055 and 3.000131), four trajectories with negligible position and velocity errors have been identified. 
Such errors represent a state mismatch at the Poincar\'e section between the stable and the unstable portion, but the magnitude is negligible (well below 1 km in position and less than 1 m/s in velocity) and can be dealt with during the navigation. Hence, these solutions can be regarded as maneuver free.
They correspond to Jacobi constant values in the middle of the assumed range, hence they are associated with Halo orbits of intermediate size. 
Although a refined search could widen the result set around these orbits, they can be considered representative and illustrative of the behaviour and performance of this type of solutions. On the other hand, it must be emphasized that HIM trajectories associated with large or small Halo orbits exhibit a large rate of collision with Enceladus or escape from the Hill sphere and are less likely to provide maneuver-free connections. The solutions of this work are characterized by times of flight in the range from 38 to 58 hours, large fractions of which are spent looping around Enceladus. The LPOs serve as departure and arrival gates for each transfer. Given their periodic character, they can be used as parking orbits between consecutive flights around the moon.

The inspection of the evolution of the osculating Keplerian elements has shown large variations of, in particular, the eccentricity and the inclination. At times, the former reaches escape values, which suggests the need for an appropriate navigation strategy, as expected for a chaotic system like the CR3BP and, even more so, in the light of the above mentioned escapes and collisions. The observed variations in inclination help achieve the objective of the work, i.e., the design of orbits with a significant out-of-plane motion and access to the polar regions of the moon. The distances from the surface vary between 150 and 1000 km. The speeds relative to an Enceladus-centered inertial frame are in the range 0.08 to 150 m/s, in good agreement with the reference values for Keplerian circular orbits in the same altitude interval, implying that the innovative aspect is not the low speed itself (which is however extremely convenient in the framework of an {\it in situ} mission), but the fact that these trajectories take into account the main perturbation acting on two-body orbits, i.e., the gravity of Saturn. In other words, the trajectories have been obtained in a dynamical model that can be considered realistic, hence accurate.
An aspect that certainly deserves attention is the perturbation caused by the main harmonics of the gravity field of the two primaries, particularly the $J_2$ zonal term. This is the subject of on-going work.    

The analysis of the observational performance of the proposed trajectories has shown that the complete surface of Enceladus is visible from the S/C and that uninterrupted windows of access to the southern polar region exist and extend over several hours, the specific duration depending on altitude: for example, in the s-heteroclinic transfer from the Northern Halo around $L_1$ to the Northern Halo around $L_2$ (type A) at $C_J$ = 3.000118 the south pole is seen during two hours from below 400 km altitude, whereas in the solution connecting a Northern Halo  with a Southern Halo at $L_1$ (type C) with $C_J$ = 3.000072 the south pole is visible for over 20 hours distributed along four windows at different altitudes.
The detailed assessment of the time of overflight (defined as the time spent by the S/C above the local horizon) of a regular grid of points over the surface has been expressed in the form of geographical color maps. These maps show that the local cumulative visibility is never shorter than 4 hours (polar regions) with peaks of 20 or even 40 hours for wide equatorial bands (up to $\pm$ 60 degrees latitude).  
Eventually, the ground tracks help understand the motion of the S/C with respect to the surface. This motion (computed taking into account the spin rate of the moon) exhibits both prograde and retrograde components, a fact that reflects the large variations in the osculating Keplerian elements in this dynamical model. 

In conclusion, the trajectories designed and studied in this investigation exhibit appealing properties that make them suitable science orbits for a future mission aiming at giving answers to our fundamental questions regarding the origin and nature of the peculiar features detected at Enceladus.

\section*{Acknowledgements}
This work been supported by Khalifa University of Science and Technology's internal grants FSU-2018-07 and CIRA-2018-085.  

\bibliography{mybibfile}

\end{document}